\documentclass{article}
\usepackage{graphicx}

\usepackage{amsmath,amssymb,amsfonts,epsfig,authblk}

\newtheorem{prop}{Proposition}

\newtheorem{cor}[prop]{Corollary}
\newtheorem{rem}[prop]{Remark}
\newtheorem{lem}[prop]{Lemma}

\newcommand{\remnn}{\medskip \noindent {\bf Remark }}

\newcommand{\proof}[1]{\vspace{1.5ex}\noindent{{\bf Proof:} #1 
}
\vspace{1.5ex}
}

\renewcommand{\theequation}{\arabic{section}.\arabic{equation}}

\newcommand{\beqa}{\begin{eqnarray*}}
\newcommand{\eeqa}{\end{eqnarray*}}

\newcommand{\beqan}{\begin{eqnarray}}
\newcommand{\eeqan}{\end{eqnarray}}

\newcommand{\beq}{\begin{equation}}
\newcommand{\eeq}{\end{equation}}

\newcommand{\ts}[1]{\protect{{\textstyle{ #1}}}}
\newcommand{\tsf}[2]{\protect{{\textstyle{ \frac{#1}{#2}}}}}

\newcommand{\mc}[1]{\protect{{\mathcal{#1}}}}
\newcommand{\mf}[1]{\protect{{\mathfrak{#1}}}}

\newcommand{\ul}[1]{\protect{{\underline{#1}}}}

\newcommand{\id}{\protect{\rm{id}}}
\newcommand{\ima}{\protect{\rm{im}}}
\newcommand{\End}{\protect{\rm{End}}}
\newcommand{\diag}{\protect{\rm{diag}}}
\newcommand{\spann}{\protect{\rm{span}}}
\newcommand{\rmx}{\protect{\rm{x}}}

\newcommand{\C}{\mathbb C}

\newcommand{\R}{\mathbb R}
\newcommand{\Sp}{\mathbb S}

\title{String Quantization and the Shuffle Hopf Algebra}

\author[1]{Dorothea Bahns}
\author[2]{Jasmin D. A. Meinecke}

 \affil[1]{{\small Courant Research Centre `Higher Order Structures in Mathematics', University of G\"ottingen,  \vspace{-0.7ex} \newline 
Bunsenstr. 3-5, 
D - 37073 G\"ottingen, Germany --- bahns@uni-math.gwdg.de}}

\affil[2]{{\small Centre for Quantum Photonics, H.H. Wills Physics Laboratory, 
University of Bristol,  
 \vspace{-0.7ex} \newline 
Tyndall Avenue,  Bristol BS8 1TL, UK  --- jasmin.meinecke@bristol.ac.uk}}

\date{}

\begin{document}

\maketitle

\begin{abstract}
\noindent The Poisson algebra $\mf h$ of invariants of the Nambu-Goto string, which was first introduced by K. Pohlmeyer in 1982, is described using the Shuffle Hopf algebra. In particular, an underlying auxiliary Lie algebra is reformulated in terms of the image of the first Eulerian idempotent of the Shuffle Hopf algebra. This facilitates 
the comparison of different approaches to the quantization of $\mf h$.

\end{abstract}


\section{Introduction}

Originally intended as a model to describe meson physics in the end 1960's, 
the action functional of the Nambu-Goto string is the straightforward generalization of the action functional of a relativistic point particle that freely moves in Minkowski spacetime $(\R^d,\eta)$ with its pseudo-metric $\eta$,
to an action functional describing  a freely moving 1-dimensionally extended object (a string). As such, instead of measuring the length of a curve, it measures the area of a surface immersed in $(\R^d,\eta)$ with respect to the induced metric. A stationary point of this functional is a surface of extremal area, called a worldsheet in analogy with the worldline of a particle  -- and in fact, endowing the embedding space $\R^d$ with the Euclidean metric  in place  of $\eta$, the stationary points of the corresponding action functional are minimal surfaces. 

Using methods of integrable systems, an infinite dimensional Poisson algebra was explicitly constructed in terms of functionals on a worldsheet which are invariant under changes of the surface's parametrization~\cite{pm_grp}. It was furthermore shown that, at least for certain classes of surfaces, the worldsheet can be reconstructed from the elements of this Poisson algebra~\cite{pr_complete}.

In this paper, we are concerned only with this Poisson algebra and its quantization. For the sake of completeness  we now give an indication of the geometric origin of these algebraic structures, but the following remarks are not essential to understand the paper's content. 
Let an immersed surface be parametrized by a map $x: \Sp^1\times \R \rightarrow \R^d$, where for each $\tau  \in I:= (\tau_0,\tau_1) \subseteq \R$, the map $\gamma_\tau: \Sp^1 \rightarrow \R^d$, $\gamma_\tau(\sigma)=x(\sigma, \tau)$ describes a   
spacelike closed curve.
Fix $\tau \in I$ and consider the following  iterated integrals for $n \in \mathbb N$,
\beq\label{eq:Rpm}
\mc R^\pm_{a_1\cdots a_n} (\sigma, \tau) := \int_{\sigma \leq \sigma_n \leq \dots \leq \sigma_1\leq \sigma +2 \pi} u^\pm_{a_1}(\sigma_1,\tau) \cdots 
 u^\pm_{a_n}(\sigma_n,\tau) \ d\sigma_1 \cdots d \sigma_n  
\eeq
where $u^\pm_{a}(\sigma,\tau)$ for $a \in \{0,\dots, d-1\}$, is the $a^{\rm th}$ component of the tangent vectors 
$u^\pm(\sigma,\tau)=\partial_\sigma x(\sigma, \tau) \pm \partial_\tau x(\sigma, \tau)$. Now, cyclically symmetrized linear combinations of these iterated integrals,
\beq\label{eq:Zpm}
\mc Z^\pm_{a_1\cdots a_n} (\sigma, \tau) := 
\mc R^\pm_{a_1\cdots a_n} (\sigma, \tau) + 
\mc R^\pm_{a_n a_1\cdots a_{n-1}} (\sigma, \tau) + \dots +  
\mc R^\pm_{a_2\cdots a_n a_1} (\sigma, \tau) 
\eeq
turn out to be independent of the starting point of the integration $\sigma$. Moreover, they are invariant under general reparametrizations of the circle, ${\rm Diff}(\Sp^1)$. And finally, if $x$ parametrizes a  worldsheet (`on-shell' case), $\mc Z^\pm_{a_1\cdots a_n} (\sigma, \tau)$ is also independent of $\tau$. 
This justifies to call them the {\em invariant charges} of the Nambu-Goto string.

As a set of iterated integrals, the linear span of $\mc R^\pm$ is endowed with the shuffle multiplication $\#$, and it was shown in~\cite{pr_algprop} that the span of invariant charges, $\mf h$, is closed under this multiplication. Moreover, a formula for a Poisson bracket of invariant charges was given there, which is derived from an antisymmetric bilinear map given in terms of the vector components of tangent vectors $u^\pm$ that generalizes the canonical Poisson bracket of a system of classical mechanics. With this bracket  and the shuffle multiplication, the span of invariant charges, $\mf h$, is a Poisson algebra.

Now, the linear span of the expressions $\mc Z^+$ Poisson-commutes with that of the $\mc Z^-$, and their structure constants differ only by a global minus sign, so one usually only considers one of these two sectors and denotes it also by $\mf h$. The deformation of the resulting Poisson algebra $(\mf h, \#, \{ \})$ has been the subject of almost three decades of research. The big picture behind this is that such a deformation yields a {\em quantization} of a geometric object which, by construction, respects its {\em invariance under reparametrizations}. 

In contrast to this, the ordinary approach to the quantization of strings  is based on methods from conformal field theory and the system's invariance under reparametrizations has to be restored by hand after the quantization. While this requires the dimension $d$ of the embedding vector space to be fixed to a particular value (the `critical dimension', $d=26$ for the Nambu-Goto string, $d=10$ in supersymmetric theories), no such need to fix $d$ has arisen so far in the framework of quantizing the Poisson algebra $\mf h$ -- hence, no extra dimensions have to be postulated there. Moreover, it was shown in~\cite{bahns_CQ} that the methods from conformal field theory do not yield a consistent quantization of the algebra of invariants in {\em any} dimension. 
So the two approaches to string quantization  have to be seen as being mutually exclusive. 
Of course, in the long run, for a physically meaningful quantum theory, one also has to understand the representation theory of a quantization of $\mf h$ -- and this is a difficult independent task, while in the conformal field theory framework, quantization automatically comes with a representation (Fock space). 

In this paper, we will reformulate the algebraic structure of $\mf h$ and the existing proposals for its quantization in a language which is well-known in combinatorial algebra. Main ingredient is the  observation that the algebraic properties of the invariant charges are captured by the letters $a_i \in \{0,1,\dots,d-1\}$ alone. We therefore make the identification 
\beqa
\mc Z^+_{a_1 \cdots a_n} &\leftrightarrow &\mbox{cyclic sum of words} 
\\&& a_1 a_2 \cdots a_{n-1} a_n +  a_n a_1 \cdots a_{n-2} a_{n-1} +  \dots
+  a_2  a_{3} \cdots a_n a_1
\eeqa
We will then be able to understand the set of so-called {\em truncated tensors} which play an important role in the structural investigation of $\mf h$, as the image of the first Eulerian idempotent in the word algebra, and thereby considerably simplify a number of proofs. This is the main point of the following section. In the third section of this paper, we will then explain different approaches to the quantization of $\mf h$, especially the proposal of Meusburger and Rehren~\cite{meusb_diplom, rm_real}. This approach 
yields a quantization of the Poisson algebra $\mf h$, {\em provided} that a certain conjecture on the structure of $\mf h$, the so-called  {\em quadratic generation hypothesis}, turns out to be true. We will understand this approach within a more general setting  of deformations and point out why it seems to be impossible to extend it in such a way as to make the quadratic generation hypothesis unnecessary. In the outlook we will briefly contrast this approach with a more recent proposal~\cite{beh_pm}  to quantize $\mf h$ as a Quasi-Lie-bialgebra. Part of the results presented have also been worked out in the {\em Diplom} thesis~\cite{meinecke_diplom}. 


\section{The auxiliary Lie algebra}\label{sec:euler}
\setcounter{equation}{0}

We first recount some facts about Hopf algebras. In a Hopf algebra $H$ over a field $k$ (of characteristic 0), we denote the multiplication map by $\mu : H \otimes H \rightarrow H$, the unit by $\i: k\rightarrow H$, the co-multiplication map by $\Delta:H \rightarrow H \otimes H$, the co-unit by $\epsilon:H\rightarrow k$, and the antipode by $S:H \rightarrow H$.  Observe that in Hopf algebras the multiplication is always assumed to be associative and the co-multiplication is assumed to be co-associative.

Let $f, g:H\rightarrow H $ be linear, then the convolution  $f*g$ of $f$ and $g$ is the linear map
\[
f*g=\mu \circ(f \otimes g)\circ \Delta \ : \ H\rightarrow H
\]
Taking the convolution is an associative operation and $\i\eta:= \i\circ \eta:H\rightarrow H $ is the neutral element which turns $(\End_k \, H,+,*)$ into a unital algebra. 
Now, let $H$ be  a commutative graded connected Hopf algebra, $H= \bigoplus_{n\geq 0} H_n$ with $H_0=k$.  Consider a linear map $f \in \End_k H $ with $f(1)=0$, and assign to it a linear map $l(f) \in \End_k \, H$ defined by
\[
l(f):= \ln_*(\i\eta+f) =  f - \tsf 1 2 \, f^{*2}  + \tsf 1 3 \, f^{*3} + \dots + \tsf{(-1)^{j+1}} j \, f^{*j} + \dots  
\]
The sum above is in fact finite when applied to any $x \in H$, since for $f$ with $f(1)=0$, we have  $f^{*k}|_{H_n}=0$ for $k>n$. Similarly, one defines linear maps  $l^{(k)}(f)$ by
\[
l^{(k)}(f):= l(f)^{*k}  / k!
\] 
with the convention 
$l^{(0)}(f)=\i\eta$, and consistent with $l^{(1)}(f)=l(f)$. 

Now consider the map $f= \id -\i\eta$. Obviously, it satisfies $f(1)=0$, since 
$\i\eta(1)=1$.
The resulting maps $e^{(k)}:=l^{(k)}(\id -\i\eta):H \rightarrow H$ have the following properties:

{\prop \label{prop:e_projId} \rm Let $H$ be a commutative graded connected Hopf algebra. Then for 
 $e^{(k)}:=l^{(k)}(\id -\i\eta):H \rightarrow H$ we have
\beq
\label{eq:e_Id} \mbox{$\id|_{H_n} = e^{(1)}|_{H_n} + \dots + e^{(n)}|_{H_n}$ for $n \geq 1$ \phantom{and}}
\eeq
and
\beq
\label{eq:e_proj} \mbox{$e^{(k)}\circ e^{(k)}=e^{(k)}$ and $e^{(k)}\circ e^{(j)}=0$ for $k \neq j$} . 
\eeq
}

In the literature,  $e^{(k)}$ for $k=0,1,\dots$, is usually called the $k$-th Eulerian idempotent\footnote{Usually, in the literature, also $l$ is denoted by $e$. We have slightly changed this convention here in order to avoid confusion.}, which is justified by the second property above. Moreover, in this paper, we will call the elements of $\ima \, e \subset H$ the Euler elements in $H$.

A proof of the above proposition can be found in e.g.~\cite{lod_idem} or \cite[Sect 4.5.2]{lod_book}. It relies on the fact that by the identity $(1+\rmx )^p=\exp(p \ln(1+\rmx))$, one has $(\i\eta +f)^{*p}=\i\eta + \sum_{j\geq 1}p^j \, l^{(j)}(f)$ for $p\geq 1$,  so it follows, in particular, that
\beq\label{eq:resol}
\id^{*p}|_{H_n} =  \sum_{j=1}^n p^j\,  e^{(j)}|_{H_n} \qquad \mbox{ for } n\geq 1, p\geq 1\ .
\eeq

Also from the identity (\ref{eq:resol}), together with the fact that $\id^{*p}$ is an algebra-homomorphism, we find that in a commutative, graded connected Hopf algebra, $e$ vanishes on a product $X_1 X_2$ of  algebra elements $X_i \notin H_0$,  
\beq\label{eq:emult0}
e (X_1 X_2) =0
\eeq
To see this, apply (\ref{eq:resol}) to both sides of 
$\id^{*p}(X_1 X_2)=\id^{*p}(X_1) \,  \id^{*p}(X_2)$.

By the same argument, applied to an $L$-fold product $X_1 \cdots X_L$ of algebra elements $X_i \notin H_0$, it follows, more generally, that 
\[
e^{(K)}(X_1 \cdots X_L) =  \left\{ \begin{array}{ll}
0 & \mbox{for } K<L \\
\sum_{\sum K_i=K, \;  K_i \geq 1} e^{(K_1)}(X_1)\cdots e^{(K_L)}(X_L)& \mbox{for }  K\geq L
\end{array}\right.
\]
In particular, for $X_i=e(Y_i)$, we have $e^{(K)}(e(Y_1) \cdots e(Y_K)) = e(Y_1)\cdots e (Y_K)$. 
We deduce, in particular, that if any algebraic dependences between elements of $\ima \, e$ persist, they have to be homogeneous: Let $P$ be a finite sum of the form
\[
P=\sum_{K\geq 0 ,X_i \in H} c_{X_1, \dots , X_K} e(X_1) \cdots  e(X_K)
\]
and suppose $P=0$. Let $K_0$ be the smallest of the $K$'s which occur in this expression. Apply $e^{(K_0)}$ to $P$, then all contributions with $K$ larger than $K_0$ are mapped to 0, while the contribution with $K=K_0$ remains unchanged. It follows that this contribution has to be 0 in itself. Repeating this argument for increasing $K$ yields the claim. It is therefore justified to think of $K$ as a polynomial degree.

It follows that we can use the decomposition (\ref{eq:e_Id})
to define a new grading (`polynomial degree', or following~\cite{pr_algprop}, `homogeneity degree') in addition to the original one in $H$, by
\beq\label{eq:homdeg}
H=\bigoplus_{K\geq 0} H^{(K)} \quad \mbox{with } H^{(0)}=k\, , \  H^{(K)}= \spann_k\{e(X_1)\cdots e(X_K) \; | \; 
X_j  \in H \}
\eeq
Observe that in the definition of $H^{(K)}$ for $K\geq 1$, we did  not have to specify that $X_i\notin H_0$, since $e(1)$ is $0$. By construction, the algebra multiplication is of degree 0 with respect to this grading.

Let us now consider the particular case of the Shuffle Hopf algebra. Let $H$ denote the free module over a field $k$ of characteristic 0, with basis given in terms of all words from an alphabet $A$ of $d$ letters (i.e. $H \cong T(V)$ with $V$ a $d$-dimensional vectorspace over $k$). As a vectorspace, $H$ is graded with respect to the word-length, and  $H_0$ is identified with the groundfield $k$.  We will generally use $x_i,y_j,\dots$ to denote letters from the alphabet $A$, and $x,y,\dots$ as well as $X_i, Y_j, ...$ to denote words. We now equip $H$ with the shuffle multiplication, which for $x=x_1\cdots x_n \in H_n$ and $y=y_1\cdots y_k\in H_k$ is given by
\[
x \# y := \mu(x \otimes y) := \sum_{\sigma \in S_{n+k,n}} \sigma (x_1\cdots x_n y_1 \cdots y_k)
\]
where $ S_{n+k,n} \cong S_{n+k}/S_n \times S_k$ is the set of permutations which do not change the order of the first $n$ and the last $k$ letters (`Shuffle' permutations\footnote{The name is appropriate because the action of such a permutation corresponds to taking two decks of cards and once shuffling one of the decks into the other.}), and where the natural action of the permutation group on a word is given by $\sigma(x_1\cdots x_n)=x_{\sigma(1)}\cdots x_{\sigma(n)}$. An example is $ab \# c = abc+acb+cab$ where  $a,b,c \in A$.
The shuffle multiplication is obviously graded with respect to the word length, it is commutative, and the neutral element is the empty word $\emptyset=1 \in k$. 
 
It is often convenient to use one of the following, equivalent, recursive definitions of the shuffle product,
\beq\label{eq:recshuff}
x_1 \cdots x_n \# y_1 \cdots y_k 
\ \begin{array}[t]{l}= \  x_1 \cdot (x_2 \cdots x_n \# y_1 \cdots y_k) +  y_1 \cdot (x_1 \cdots x_n \# y_2 \cdots y_k) 
\\
= \ (x_1 \cdots x_{n-1} \# y_1 \cdots y_k)\cdot x_n +   (x_1 \cdots x_n \# y_1 \cdots y_{k-1})\cdot y_k 
\end{array}\eeq
where $x_i, y_j  \in A$, and $\cdot$ denotes the concatenation product.

It is well-known that $(H,\#)$ becomes a Hopf algebra when equipped with the deconcatenation co-product given by
\[
\Delta(x_1\cdots x_n) = 1 \otimes x_1\cdots x_n + x_1 \otimes x_2 \cdots x_n + \dots 
+ x_1\cdots x_{n-1}\otimes x_n + x_1\cdots x_n \otimes 1
\]
with co-unit $\eta(x)=0$ unless $x=1$, and with the antipode 
$
S(x_1\cdots x_n)=(-1)^n x_n x_{n-1}\cdots x_2 x_1  \ .
$
Now observe that in the Shuffle Hopf algebra, we have $(\id - \i \eta)\otimes (\id - \i \eta) (x\otimes y)=x\otimes y$ for $x,y \neq 1$ (and, as usual,  0 if one of them is the empty word). It follows that the first Eulerian idempotent is given explicitly by
\beq\label{eq:eofx}
e(x_1\cdots x_n)=\sum_{k=1}^n \frac {(-1)^{k+1}}{k} \sum_{I_1\sqcup \dots \sqcup I_k=\ul n} x_{I_1}\# \cdots \# \, x_{I_k} 
\eeq
for any word $x_1\cdots x_n \in H_n$, $n\geq 1$.
Here, the second sum runs over all ordered partitions of the ordered set $\ul n :=\{1,\dots n\}$ into {\em non-empty} sets $I_j$ (i.e. for $\{1,2,3\}$ and $k=2$, we consider $I_1=\{1\}, I_2=\{2,3\}$ and $I_1=\{1,2\}, I_2=\{3\}$). For an ordered index set $I=\{i_1,\dots,i_s\}$, we denote the word $x_{i_1} \cdots x_{i_s}$ by $x_I$.
Recall here that $e(1)=0$.

In a similar way, the first identity (\ref{eq:e_Id}) from Proposition~\ref{prop:e_projId} yields the decomposition
\beq\label{eq:xdecEul}
x_1\cdots x_n=\sum_{k=1}^n \frac 1 {k!}\; \sum_{I_1\sqcup \dots \sqcup I_k= \ul n}e(x_{I_1})\# \dots \# e(x_{I_k})
\eeq
for any word $x_1\cdots x_n \in H_n$, $n\geq 1$.

These two identities, as well as (\ref{eq:emult0}) for the shuffle multiplication, were proved in~\cite{pr_algprop} (Propositions 2 and 3), where the elements of $\ima\, e$ (called the `truncated tensors')  were constructed explicitly from the logarithm of a monodromy matrix and the proofs were not directly based on equation (\ref{eq:resol}).  
In this explicit framework, the authors then investigated further algebraic structures, most importantly, a Lie bracket on the image of the first Eulerian idempotent. Contrary to the general constructions  discussed above, these algebraic structures seem to be defined only in the Shuffle Hopf algebra. 

We start with proving some helpful identities. 

{\lem {\rm The first Eulerian idempotent of the Shuffle Hopf algebra satisfies
\beqan
e(x_1\cdots x_n)&=&
(-1)^{i-1}\, e\big(x_i\cdot (x_{i-1} \cdots x_1 \# x_{i+1} \cdots x_n)\big)
\label{eq:pm7a}\\
&=&
(-1)^{n+i}\, e\big((x_1 \cdots x_{i-1} \# x_n \cdots x_{i+1})\cdot x_i\big)
\label{eq:pm7b}
\eeqan
for $i\in\{1,\dots, n\}$. }}

Here, as well as in what follows, we adopt the convention that an empty string, e.g. $x_{i-1}\cdots x_1$ for $i=1$ is   understood to be 1.

{\proof{We follow the proof from \cite{pr_algprop}. From the recursive definition of the shuffle product (\ref{eq:recshuff}), we deduce that for $i\in \{0,\dots, n\}$,
\beqan
&&x_i \, (x_{i-1}x_{i-2}\cdots x_1 \# x_{i+1}\cdots x_n)
\nonumber \\&& \quad = x_i x_{i-1}\cdots x_1 \# x_{i+1}\cdots x_n - x_{i+1}\,(x_i x_{i-1} \cdots  x_1 \# x_{i+2} \cdots x_n) \qquad \qquad 
\label{eq:recder}
\eeqan
where we suppress the concatenation product in the notation. Using also (\ref{eq:emult0}), i.e. 
$e(x\# y)=0$ for $x,y\neq 1$, we then find, e.g.
\beqa
e(x_1\cdots x_n)&\stackrel{(\ref{eq:recder})}{=}&e(x_1\# x_2\cdots x_n) - e(x_2\, (x_1 \# x_3 \cdots x_n)) \stackrel{(\ref{eq:emult0})}{=}
 - e(x_2\, (x_1 \# x_3 \cdots x_n)) 
\\&\stackrel{(\ref{eq:recder})}{=}& - e(x_2x_1 \# x_3 \cdots x_n) + e(x_3\,(x_2x_1\#x_4\cdots x_n) )
\\ &=& \dots
\quad =\  (-1)^{n-1}\,e(x_n \cdots x_1)
\eeqa
Thus, (\ref{eq:pm7a}) is proved, from which also  (\ref{eq:pm7b}) directly follows using the last identity above, $e(x_1\cdots x_n)=(-1)^{n-1}\,e(x_n \cdots x_1)$ and the commutativity of $\#$. \hfill $\square$

}}

In~\cite{pm_solconstr}, a set $L$ of words was identified, such that $e(L)=\{e(x)\ | \ x\in L\}$ is a basis of $\ima \, e$, and an algorithm was given of how any element of $\ima \, e$ can be rewritten as a linear combination of elements from $e(L)$. It was also proved that (together with the empty word), $e(L)$ yields a generating set which {\em freely} generates the Shuffle algebra by the decomposition formula (\ref{eq:xdecEul}). 
This set $L$ consists of all words which are lexicographically strictly minimal among all cyclic permutations of their letters, and were called `cyclically minimal' in \cite{pm_solconstr}. Examples of such words are $abc$, $acb$, $aaab$, but not $abab$ or $bac$. In combinatorial algebra, such words are today known as Lyndon words, and Pohlmeyer's results from~\cite{pm_solconstr} can be understood\footnote{It seems that the results from~\cite{pm_solconstr} and~\cite{reut_book} were found independently at around the same time.
 } from Reutenauer's theorem that the Lyndon words {\em freely} generate the Shuffle algebra, cf.~\cite{reut_book}.
Observe that the number of Lyndon words of length $n$ over an alphabet of $d$ letters is  
\beq\label{eq:moeb}
\frac{1}{n} \sum_{s|n} \mu(s)\; d^{{n}/{s}}
\eeq
where the sum runs over all divisors of $n$ and where $\mu$ denotes the M\"obius function. Notice that also this formula already appeared in~\cite{pm_solconstr}. 

Let us now turn to the definition of a Lie bracket on $\ima \, e$, first given in~\cite{pr_algprop} and called the `modified Poisson bracket' there. In order to write it in terms of our Hopf algebraic language, we first introduce linear maps $\partial^L_a, \partial^R_a : H \rightarrow H$, $a \in A$,  given by
\[
\partial^L_a(x_1\cdots x_n)= \delta_{a,x_1}\,x_2\cdots x_{n} \quad \mbox{ and }  \quad 
\partial^R_a(x_1\cdots x_n)= \delta_{a,x_n}\,x_1\cdots x_{n-1}
\]
with the Kronecker-Symbol $\delta$. It is not difficult to see that these maps are derivations of the shuffle algebra, so we may interpret $\partial^L_a, \partial^R_a$ as partial derivatives.
Observe also that 
\beq\label{eq:partRL}
\partial^R_a = - S \circ  \partial^L_a \circ S 
\eeq
and that, for for $x=x_1\cdots x_n \in H_n$, we have
\beqan
\partial^R_a\otimes S (\Delta(x)) &=& 
\sum\limits_{i=1}^n (-1)^{n-i} \delta_{a,x_i} x_1\cdots x_{i-1}\otimes x_n\cdots x_{i+1}
\label{eq:partRS}
\\
S\otimes \partial^L_a (\Delta(x)) 
&=&
\sum\limits_{i=1}^{n} (-1)^{i-1} \delta_{a,x_i} x_{i-1}\cdots x_1\otimes x_{i+1} \cdots x_n
\label{eq:partLS}
\eeqan

{\prop \label{prop:Lie} \rm{Let $H$ be the Shuffle Hopf algebra for an alphabet of $d$ letters over a field $k$ of characteristic 0. Let $g$ be a symmetric $d\times d$ matrix over $k$. Then the bilinear map $[\cdot,\cdot]: \ima\, e \times \ima \, e \rightarrow \ima \, e$ given by
\beq\label{eq:Lie}
\, [e(x),e(y)] = 
\sum_{a,b \in A} g_{ab}\; e\left( \partial^R_a * S (x) \cdot S * \partial^L_b (y) \right)
\eeq
for words $x, y \in H$ with $|x|, |y|\geq 2$, and 0 otherwise, defines a Lie bracket on $\ima\, e$.
}}

{\proof{We first 
note that the definition is consistent with (\ref{eq:emult0}) since 
\[
\partial_a^R * S (x \# x^\prime) = 0  \qquad \mbox{for } x, x^\prime \neq 1
\]
and likewise for $S* \partial^L_b$. To see this, observe first that of course, we have $\Delta(x\# y) = \Delta(x)\#\Delta(y)$, such that  the contribution to $\partial_a^R * S (x \# y) $ which contains e.g. $\delta_{a,x_i}$ is
\[
\delta_{a,x_i}\ x_1\cdots x_{i-1}\# \Big( \sum_{j=0}^k y_1\cdots y_j \# S(y_{j+1} \cdots y_k) \Big) \# S(x_{i+1}\cdots x_{n})
\]
The sum in this expression is equal to
\beqa
&&(-1)^k (\ y_k\cdots y_1 \ - \  y_1 \# y_k \cdots y_2\  + \  y_1 y_2 \# y_k\cdots y_3 \  - \dots 
\\&& \qquad \dots \ -(-1)^k y_1\cdots y_{k-1}\# y_k \ + \ (-1)^k y_1\cdots y_k\ )
\eeqa
By the recursive definition of the shuffle product, the first two terms inside the bracket add up to  $-(y_1 \# y_k \cdots y_3) y_2$, which, added to the third term is equal to  $(y_1y_2 \# y_k\cdots y_4)y_3$, and so on, until we produce $-(-1)^k (y_1 \cdots y_{k-1} \#1)y_k$ for the sum up until the second before last term. Added to the last remaining term $(-1)^k y_1\cdots y_k$ this gives $0$.

To prove that the bracket is antisymmetric, observe that 
\beqa
&& e\left( \partial_a^R * S (x) \cdot S * \partial_b^L (y) \right) = 
- e\left( S\left(\partial_a^R * S (x) \cdot S * \partial_b^L  (y) \right) \right) 
\\&& \qquad  = 
- e\left( S(S * \partial_b^L (y)) \cdot S(\partial_a^R * S (x)) ) \right) = 
- e\left(\partial_b^R * S (y) \cdot  S * \partial_a^L (x) \right)
\eeqa
where the last step directly follows from  (\ref{eq:partRS}) and (\ref{eq:partLS}). Now the bracket's antisymmetry follows since $g$ is symmetric.

We now bring the bracket into a more explicit form, which was in fact the one used in \cite{pr_algprop}. We will then be able to mimic the idea of the proof of Jacobi's identity from there. By (\ref{eq:partRS}) and (\ref{eq:partLS}), the bracket of two words  $x=x_1\cdots x_n \in H_n$ and $y=y_1\cdots y_k \in H_k$ with $n, k\geq 2$ can be written as
\[
-\sum_{i=1}^n\sum_{j=1}^k g^{x_i y_j}\; (-1)^{n-i-j}\ 
e\left( (x_1 \cdots x_{i-1} \# x_n \cdots x_{i+1}) \cdot (y_{j-1}\cdots y_1\# y_{j+1} \cdots y_k)
 \right)
\]

Observe that in this explicit presentation, the bracket's antisymmetry can be proved using $e(u_1\cdots u_{n+k-2})=(-1)^{n+k-1}e(u_{n+k-2} \cdots u_1)$ (which is implied by (\ref{eq:pm7a})) 
and the commutativity of the shuffle product.

Now consider the cyclic sum
\[
\,[ e(x), [e(y),e(z)]] + [ e(z), [e(x),e(y)]] + [ e(y), [e(z),e(x)]] 
\]
Let $|x|=n, |y|=k$, and $|z|=s$. We first consider the contribution to the first term which contains the product $g_{x_n,y_1}\,g_{y_k,z_1}$,
\[
g_{x_n,y_1}\,g_{y_k,z_1}\,e(x_1\cdots x_{n-1} y_2 \cdots y_{k-1}z_2\cdots z_s)
\]
The only other such contribution can appear in the second term, rewritten as $-[[e(x),e(y)],e(z)]$, such that the term in question now appears with a minus sign. 
Similarly, the contribution to the second term which contains $g_{z_s,x_1}\,g_{x_n,y_1}$ is
\[
g_{z_s,x_1}\,g_{x_n,y_1}\,e(z_1\cdots z_{s-1} x_2\cdots x_{n-1} y_2 \cdots y_{k})
\]
and the only other such contribution  can appear in the third term, rewritten as $-[[e(z),e(x)],e(y)]$, hence as in the case before, the term appears with a minus sign. By application of (\ref{eq:pm7a}) and (\ref{eq:pm7b}), it follows that the other contributions to the cyclic sum also cancel each other.  \hfill  $\square$

}}

The Lie algebra $\mf g:=\ima\, e$ with the bracket given in proposition~\ref{prop:Lie} above, 
is a {\em graded Lie algebra} with respect to the grading given by the word length subtracted by two, 
\[
\mf g = \bigoplus_{r=-1}^\infty \mf g_{r} \qquad \mbox{ where} \quad r(e(x))=|x|-2 
\]
Observe that it is consistent that the grading starts with $-1$ since all elements in 
$\mf g_{-1}=\spann_k\,\{e(a) | a\in A\}$ are central. Observe also that each stratum $\mf g_r$ is a finite dimensional vectorspace with dimension given by the number of Lyndon words of length $r-2$ over an alphabet of length $d$, cf. (\ref{eq:moeb}).

Using the decomposition (\ref{eq:xdecEul}) and the Leibniz rule with respect to the shuffle product, the bracket on $\ima\, e$ can be extended to a Poisson bracket on the Shuffle algebra,
\beq\label{eq:PoissH}
\{\cdot, \cdot\}_g : H \times H \rightarrow H
\eeq 
where the subscript $g$ indicates the dependence of the structure constants on the choice of the symmetric matrix $g$. With respect to the homogeneity degree, cf. (\ref{eq:homdeg}), 
this Poisson bracket $\{\cdot,\cdot\}_g$ is of degree -1,
\[
\{H^{(K)},  H^{(L)} \}  \subseteq H^{(K+L-1)}
\]
and with respect to the word length, it is of degree -2. 
It was therefore proposed already in \cite{pr_algprop} to introduce a combined degree $\ell$, 
with respect to which $H$ is a graded Poisson algebra,
\beqan
&&H=\bigoplus_{\ell \geq -1} H^{\ell} \qquad \mbox{where } 
H^{\ell} = \spann_k \{  H^{(K)}\cap H_n\ | \ n-K=\ell +1\} \label{eq:combgrad}
\\
&& \mbox{such that } \{H^{\ell_1},H^{\ell_2}\}\subseteq H^{\ell_1+\ell_2}
\mbox{ and } H^{\ell_1} \# H^{\ell_2} \subseteq H^{\ell_1+\ell_2+1} \nonumber
\eeqan
Again we observe that the grading may start with degree $-1$ since 
\beq\label{eq:Hm1}
H^{-1}= 
\spann_k\big( \{1\}\cup\{e^{(K)}(x_1\cdots x_K)\ | \ K\geq 1\,, \;  x_i \in A \}\big)
\eeq
so, since $e^{(K)}(x_1\cdots x_K)=e(x_1) \# \cdots \# e(x_K)/K!$, we conclude that  
$H^{-1}$ Poisson-commutes with the full Poisson algebra  $H$. 
Observe that the strata of this grading are infinite dimensional vector spaces. We shall comment on this again in section~\ref{subsec:originalPM} below.

{\prop \label{prop:hsubalg}{\rm Let $\mf h \subset H$ denote the vector subspace  of invariant charges, i.e. the subspace of cyclic linear combinations of words, $\mf h := \ima\, Z \subset H$,
with $Z:H \rightarrow H$ the cyclic symmetrization map,
\[
Z(x_1\cdots x_n)= x_1\cdots x_n + x_nx_1\cdots x_{n-1} + \dots + x_2 \cdots x_n x_1 \ .
\]
Then $\mf h$ forms a Poisson subalgebra in $(H,\{\cdot,\cdot,\}_g,\#)$. 

Explicitly, one has  
\[
Z(x_1\cdots x_n) \# Z(y_1\cdots y_m) = Z\big(  \big( Z (x_1\cdots x_n)\#y_1\cdots y_{m-1} \big) y_m\big)
\]
and for the Poisson brackets, one finds $\{Z(x_1),Z(y_1\cdots y_k)\}=0$ for $x_1,y_j\in A$, while the contribution with coefficient $g_{x_n y_1}$ to the  bracket $\{ Z(x_1\cdots x_n ),Z(y_1\cdots y_k)\}_g$ with $x_i,y_j\in A$ and $n,k\geq 2$, is
\beq\label{eq:PoisBrZ}
-Z(x_1(x_2\cdots x_{n-1} \# y_2\cdots y_{k-1})y_k) + Z(y_2(x_1\cdots x_{n-2} \# y_3\cdots y_{k})x_{n-1})
\eeq
Here, an empty expression such as $x_2\cdots x_{n-1}$ for $n=2$, is understood to be 1.}}

{\remnn By definition,  we have  $Z(x_1\cdots x_n)=Z(x_n x_1\cdots x_{n-1})=\dots=Z(x_2\cdots x_nx_1)$,  so formula (\ref{eq:PoisBrZ}) completely determines the bracket.

With respect to the combined degree,  $\mf h$ is, of course, a graded Hopf algebra whose strata $\mf h^\ell=\ker \partial \cap H^\ell$ are still infinite dimensional vectorspaces. For instance, $\mf h^1$ contains all elements of the form $Z(e^{(K)}(x_1\cdots x_{K+2}))$ for arbitrary $K\geq 2$. Observe here that for $K=1$, the only non-zero invariants  $Z(e^{(1)}(x_1\cdots x_n))$ are those where  $n= 1$.
}

\medskip{\bf Indication of the proof of proposition~\ref{prop:hsubalg}:}{

To see that $\ima Z$ forms a subalgebra in $(H,\#)$, observe that we have $\ima Z = \ker \partial$ where
\beq\label{eq:partial}
\partial:H\rightarrow H \otimes H \; , \qquad \partial(x_1\cdots x_n)=x_1 \otimes x_2 \cdots x_n - x_n\otimes x_1\cdots x_{n-1}  
\eeq
and that, using the recursive definition of the shuffle product, it is not difficult to see that $\partial$ is a derivation of the Shuffle algebra along the algebra morphism $\varphi(x)=1\otimes x$,
\beqa
\partial(x\#y)&=&\partial(x_1(x_2\cdots x_n \#y_1\cdots y_{m-1})y_m + x_1(x_2\cdots x_{n-1} \#y)x_n
\\&& \quad + y_1(x_1\cdots x_{n-1} \#y_2\cdots y_{m})x_n+ y_1(x \#y_2\cdots y_{m-1})y_m)
\\&=& \partial(x) \# (1\otimes y)+ (1\otimes x)\# \partial(y)
\eeqa

We cannot, unfortunately, use $\partial$ to prove that $\ima Z$ is closed under taking Poisson brackets, since
$\partial$ is not a derivation with respect to the bracket, as the following simple example shows ($a,b,c,d \in A$):
 \beq\label{eq:partialNoPoD}
\partial (\{ab,cd\}) = g_{bc} (a\otimes b-b\otimes a) + g_{..} \dots \neq 0 = \{\partial(ab),1\otimes cd\}+\{1\otimes ab,\partial(cd)\}
\eeq
We will come back to the construction of derivations for the Poisson algebra $\mf h$ later. 

A direct combinatorial proof of the explicit form of the bracket of two invariants would be rather involved. We therefore refer the reader to~\cite[p.614-5]{pr_algprop} for an indirect proof.  Its idea is as follows: The bracket for the invariant charges given by (\ref{eq:PoisBrZ}) was deduced from a Poisson bracket among the left movers using the integral representation~(\ref{eq:Zpm}). In the same way, an antisymmetric bilinear map could be derived for the truncated tensors (which correspond to the elements of $\ima\, e$ in the present framework). This map violates the Jacobi identity, but by construction, when it is extended to the set of  invariant charges via the Leibniz rule, it reproduces the original bracket (\ref{eq:PoisBrZ}). Therefore, one can define a bracket on the set of truncated tensors by dropping the terms which violate the Jacobi identity, whose extension to the set of invariant charges will, however,  still reproduce the original bracket (\ref{eq:PoisBrZ}). This bracket is the `modified Poisson bracket' (\ref{eq:Lie})  we investigated here. 

Also in~\cite{pr_algprop}, a proof of the explicit formula for the multiplication can be found, which relies on the iterated integral representation.
\hfill $\square$

}


\section{Quantization}
\setcounter{equation}{0}

An algebraic quantization of the string model is given by a deformation of the Poisson algebra $\mf h$. Different approaches have been proposed. We begin this section by recalling some general facts about algebra deformations and the quantization of Poisson algebras.

Following~\cite{etingof}, we call a {\em deformation algebra} a topologically free $k$-algebra, i.e. a topologically free $k$-module $A\cong V[[h]]$ for some $k$-vectorspace $V$, which is equipped with a bilinear map $* : A\times A \rightarrow A$ that makes $A$ an associative algebra. Given an associative $k$-Algebra $A_c$, we call a deformation algebra $A$ a {\em deformation of $A_c$} if $A_c=A/hA$, in which case we can identify $A$ with $A_c[[h]]$ (as $k$-modules). If $A_c$ is commutative, a deformation $A$ of $A_c$ endows $A_c$ with a natural Poisson structure: Let  $x,y \in A_c$, choose liftings $\hat x, \hat y \in A$, then
\[
\{x,y\} := \frac 1 h (\hat x * \hat y - \hat y * \hat x) \mod h
\]
defines\footnote{The well-definedness of this definition, i.e. the independence of the choice of the liftings, follows directly from the fact that the difference of two liftings of the same element of $A_c$ must be at least of order $h$.} a Poisson bracket on $A_c$. In this case, we call $(A_c,\{,\})$ the quasiclassical limit of $A$, and $A$ a quantization of $(A_c ,\{,\})$. Observe that, given a Poisson bracket $\{,\}$ on an associative commutative algebra $A_c$, it may be impossible to find a deformation $A$ of $A_c$, such that the quasiclassical limit of $A$ is $(A_c,\{,\})$.

Often in physics, one starts off with a graded classical algebra $A_c$, and in order to understand the quantization for such a case, let us consider here a specific construction which we will encounter again below. 
Let $\mf g$ be a graded Lie algebra, then on the universal enveloping algebra $U(\mf g)$, a filtration is defined as follows: 
\beq\label{eq:filtU}
U(\mf g)_\ell := \spann \{ x_1 \cdots x_k | x_i \in \mf g\, , 0\leq \ts \sum_i \deg (x_i) + 
k -1 \leq \ell \}
\eeq
and  the degree of $X\in U(\mf g)$ is $\deg(X)=\min (\ell | X \in U(\mf g)_\ell)$.
Observe that we cannot turn this into a grading by simply using an equal sign in the definition: By permuting two neighbouring factors $x_i$ and $x_{i+1}$, we find $x_1 \cdots x_k=x_1 \cdots x_{i+1} x_i \cdots x_k + 
x_1 \cdots x_{i-1}\,z \, x_{i+2} \cdots x_k$ with $z=[x_i,x_{i+1}]$, and since
$\mf g$ is graded, we have $\deg z=\deg x_i + \deg x_{i+1}$, so the second term is of total degree 
$\deg(x_1 \cdots x_k) -1$. 
Observe that the commutator is of degree $0$ and multiplication is of degree $+1$ with respect to this filtration. 
Moreover, we have $U(\mf g)_\ell/U(\mf g)_{\ell -1}\cong S(\mf g)^\ell=:A_c^\ell$, the space of 
polynomials over $\mf g$ of degree $\ell$ corresponding to the filtration (\ref{eq:filtU}), i.e. $\ell= \sum \deg(x_i)+k-1$ for a (commutative) monomial $x_1\cdots  x_n$ in $S(\mf g)$. 
Observe that the commutative algebra $A_c=\bigoplus_\ell A_c^\ell$ (the symmetric enveloping algebra of $\mf g$) carries a Poisson structure given by the  extension of  the Lie bracket on $\mf g$ via the Leibniz rule. 

Now we modify the structure constants in $U(\mf g)$ by multiplication with a formal parameter $h$, and call the resulting algebra $U(\mf g_h)$. Then for $x,y, \in \mf g$, identified with the corresponding elements in $U(\mf g_h)$, we have
\[
[x,y]= \frac 1 h (x y -y x)
\]
where the bracket on the left hand side is the original one on $\mf g$. 


In spirit following \cite{etingof} once more, we now consider 
the following topologically free algebra
\beq\label{eq:QuantLie}
\mc A_{\mf g}:=\Big(\widehat{\bigoplus}_{\ell\geq 0}h^\ell U(\mf g_h)_\ell\Big)[[h]] := \big\{\ts \sum h^i v_i \; |\; i \geq \deg(v_i) \; , i -\deg(v_i) \stackrel{i\rightarrow \infty}{\longrightarrow} \infty \big\}
\eeq
and a map $\varphi : \mc A_{\mf g} \rightarrow A_c$, which maps $\ts \sum h^i v_i$ to the sum of the leading order contributions of each $v_i$, i.e. to the contribution with degree {\em equal} to $i$,
and projects it down to  $A_c$,
\beq\label{eq:varphi}
\varphi(\ts \sum h^i v_i)={\rm p}\circ{\rm lead_0}(v_0)+{\rm p}\circ{\rm lead_1}(v_1)+\dots\, . 
\eeq
Observe that the sum on the right hand side is finite, since, by the definition of $\mc A_{\mf g}$, we have $\deg(v_i)<i$ for almost all $i$. Since $\ker \varphi=h\mc A_{\mf g}$ and $\varphi$ is surjective, $\mc A_{\mf g}$ is a quantization of $A_c$. Observe that by construction, the Poisson bracket of $A_c$ is reproduced modulo $h$ by the commutator in $\mc A_{\mf g}$. 
Given  an element $X \in A_c^\ell$, a lifting in  $\mc A_{\mf g}$ of the form
\beq\label{eq:lifting}
(h^\ell X) + h (h^{\ell -1} Y_{\ell -1}) + \dots +  h^{\ell -1} (h Y_{1}) + h^\ell ( Y_0)  \quad  \mbox{ where } Y_{\ell^\prime} \in U(\mf g_h)_{\ell^\prime} \ ,
\eeq
is called a {\em quantum counterpart} for $X$, and the additions from lower strata are referred to as {\em quantum corrections}. By construction, $\varphi$ indeed maps this lifting to $X$, since ${\rm lead_\ell}$ maps all $Y \in U(\mf g_h)_{\ell^\prime}$ with $\ell^\prime <\ell$ to 0. Observe that we identify in the notation $X \in A_c$ with $X \in U(\mf g_h)$, since the ordering ambiguity we have in the definition of the latter can be absorbed in the quantum corrections.

It is of course well known that there are a number of difficulties in the theory of quantization: First of all, 
although, by Kontsevich's proof, a quantization of $C^\infty(M)$ does exist for the physically important case of Poisson manifolds $M$, it is also  known that  not every  Poisson structure has a quantization~\cite{M}. Secondly, it is known that symmetries of a classical system are not in general preserved by quantization.
(even for Moyal-Weyl quantization,  one finds that only linear symplectic diffeomorphisms and translations 
of $\R^{2n}$ lift to automorphisms of the quantization). 

Given a physical system with gauge symmetries, the problem becomes particularly difficult. In such a system, the physically relevant sector is in general given as a subspace of an algebra $A_c$, made up of those elements in $A_c$ which are invariant under gauge transformations. It is in general difficult, and sometimes impossible to quantize this subspace directly, so, very often, one looks for a quantization $A$ of $A_c$ instead and identifies a subalgebra of `quantum observables' in $A$. At least on a subspace of some representation space of $A$, the elements of this algebra should be liftings of classical physical observables, as a consequence of which, in particular, all quantum corrections in the lifting of an observable would  themselves be liftings of  observables {\em and} (by the subalgebra property) also all quantum corrections to Poisson brackets (and products) would be liftings of observables (at least on the physical subspace). Since the symmetry itself does not in general lift, this is a difficult task. Methods to provide a sensible quantization in certain cases exist nonetheless -- most notably perhaps the so-called BRST framework. 
In cases where the observables admit for a quantization themselves -- possibly after one has reduced the phase space by taking it modulo the action of the gauge group (`fixing the gauge') -- this will in general yield a different result (`reduction and quantization do not commute').

Also the results obtained in~\cite{bahns_CQ} should be regarded in the framework of this general set of questions. In canonical quantization of string theory, one applies the framework of conformal field theory by replacing the positive and negative Fourier modes $\alpha_a^{\pm n}$ of a parametrization of the worldsheet (at fixed time parameter) by  annihilation and creation operators on Fock space (where $a\in\{0,1,\dots,d-1\}$ labels the vector components in the embedding vectorspace $\R^{d}$). This gives a very large associative filtered algebra $A$, given by the (normally ordered) noncommutative polynomials in annihilation and creation operators, which automatically comes with a representation on  Fock space. Of course, in this setting the system's reparametrization symmetry is lost at first -- even the separation of positive and negative modes is not an invariant concept: after a reparametrization, the original positive modes cannot in general be written in terms of the new positive modes alone, but only as a mixture of positive and negative modes\footnote{This can be seen already in the example of the Fourier series of a function on the circle and a M\"obius transform.}. The physically relevant sector of $A$ and of the representation space are then identified using e.g. the BRST framework, and consistency requirements force $d$ to take a particular value (`critical dimension', $d=26$ for the Nambu-Goto string). This identification of the physical sector, however, does not seem to satisfy the requirements sketched in the previous paragraph: While the classical observables of the theory, i.e. the elements of the Poisson algebra $\mf h$, can be written in terms of the Fourier modes $\alpha_a^{\pm n}$,  their normally ordered counterparts do not form a subalgebra in $A$, not even on the physical subspace, nor in the critical dimension.

It is therefore worthwhile to pursue a different programme to string quantization. Now, originally~\cite{pm_grp}, it was proposed to directly find a quantization of the Poisson algebra $\mf h$. The philosophy behind this approach is that  the (quasi)-classical limit has a physical meaning only for observable quantities -- and that by imposing this limit  on a larger auxiliary algebra, such as $H$, one might lose information about the quantization of the physical sector of the theory. We will sketch this approach in the next subsection. It will also be explained why it was clear from the start, that the techniques employed could not lead to an actual quantization of $\mf h$. The existence of a quantization, at least under the prerequisite that a certain conjecture about the structure of $\mf h$ ('quadratic generation hypothesis') is true, was first proved in~\cite{rm_real}. Here, the quantization is explicitly realized as a subalgebra in a quantization of $H$. In section~\ref{subsec:rm}, this approach will be explained in detail employing the language developed above. Finally, as an outlook, we will briefly comment on a quantization proposed in~\cite{beh_pm} which is based on the quantization of Quasi-Lie-bialgebras~\cite{enriquez}.


In what follows, we will assume
the ground field $k$ to be the complex numbers $\C$, and the alphabet $A$ to be $\{0,1,\dots,d-1\}$, the interpretation of which is that the alphabet $A$ runs through the labels of a vector in the embedding Minkowski space $\R^d$.
More importantly, we will assume the symmetric matrix $g$ in the structure constants of $\mf g$ to be of maximal rank~$d$.


\subsection{The original programme}\label{subsec:originalPM}

Let us start by sketching the idea of Pohlmeyer's original quantization programme. The first step here is to take the fact seriously that $\mf h^{-1}=H^{-1}$ is in the algebra's centre and therefore to treat all its elements as scalars. To see what this means, we first recall that  $\mf h^{-1}$ is generated via the shuffle product by 1 and $\{e(a)|a\in A\}$, cf. the comment following~(\ref{eq:Hm1}). Now, the Poincar\'e group acts on $\mf h$,  respecting the combined  degree $\ell$, and the physical interpretation of the elements $\{e(a)|a\in A\}$, is that they are the components of the string's total momentum, $\mc P_a=e(a)$. Therefore, one has two possibilities here from the point of view of physics: Either one assumes that the Lorentz square $\mc P^2=\sum_{a,b}\mc P_a \eta_{ab} \mc P_b$ is 0 (massless case) or that it is equal to a positive parameter $m^2$, the square of the string's mass $m>0$ (massive case). In the latter case, one sets $e(0)=m$, and $e(a)=0$ for $a \in A\setminus \{0\}$ and uses $m$ as a free parameter in the theory, denoting the resulting algebra by $\mf h_m$. The physical interpretation of this particular choice is that one considers a (massive) string in its rest frame, and that due to covariant action of the Poincar\'e algebra on $\mf h$, no information is lost. Some calculations along similar lines as those described below have been performed also for the massless string, but the machinery is much less developed there. Observe that from the geometric construction of the invariants, it follows that  $g$ is proportional to
the Minkowskian pseudo-metric, $g=-2\eta=-2\, \diag (+1,-1,\dots,-1)$. Therefore, all calculations in the original Pohlmeyer programme were performed with this $g$, although this is not necessary from an algebraic point of view.

In~\cite[Prop.~17]{pr_algprop}, it was shown that $\mf h_m$ (for any choice of $m>0$) is {\em freely} generated as an algebra (with shuffle multiplication) by the (infinite) set of so-called standard invariants, 
\beqa
&&Z(e^{(2)}(0ab)) \in \mf h_m^0\, , \quad 
K^{-1} Z(e^{(2)}(0a \underbrace{0\cdots 0}_{K-1} b))\in \mf h_m^{K-1}\, , \ \mbox{ and }
\\&&
(K-1)!\, Z(e^{(K)}(\underbrace{0\cdots 0}_{K-1}\,a\, x_1 \cdots x_{K-1}\, b))
\in \mf h^{K-1}_m \, , 
\\&&
\mbox{where } K\geq 2\,,\ a,b \in A\setminus\{0\} \,,\ 
\mbox{not all } x_j \mbox{ equal } 0
\eeqa
For given $K$, the number of standard invariants is finite -- in fact, it can be explicitly calculated for any size of the alphabet $A$ using a variant of (\ref{eq:moeb}). So, in every stratum, there is only a finite number of standard invariants, although, as we pointed out before,  the strata are infinite dimensional vector spaces. From this, it  is not difficult to see that the standard invariants do not freely generate $\mf h$ as a Poisson algebra, since the number of products and Hall-basis (multiple) Poisson brackets\footnote{The Hall basis is used here since one only wants to count (multiple) brackets which are independent when the bracket's antisymmetry and the Jacobi identity are taken into account.} of standard invariants of lower strata which lie in a given stratum $\mf h_m^\ell$ is in general larger than the number of standard invariants in $\mf h_m^\ell$. It was moreover shown that there is an infinite number of linear combinations of standard invariants of homogeneity degree 2, called the exceptional elements, which are not in the linear span of Poisson brackets of standard invariants of lower strata. For a thorough discussion see~\cite{pr_algprop}. More such exceptional elements were found more recently, cf.~\cite{pm_last}. As a consequence, $\mf h_m$ is not finitely generated as a  Poisson algebra by a finite set of standard invariants. 

It is, however, conjectured that the space $\mf U$ which is generated as a Poisson algebra by the standard invariants from stratum $\mf h^0_m$ and $\mf h^1_m$, is a Poisson subalgebra of $\mf h_m$, such that $\mf h_m$ is the semi-direct product of $\mf U$ with an Abelian subalgebra $\mf a$ spanned by (modified) exceptional elements acting on $\mf U$. Again, counting the number of products and Hall basis (multiple) brackets of standard invariants from $\mf h^0_m$ and $\mf h_m^1$ in a given stratum $\mf h_m^\ell$ and comparing them to the number of standard invariants in $\mf h_m^\ell$, except those contributing to (modified)  exceptional elements, one finds that $\mf U$ cannot be freely generated as a Poisson algebra. The difference of these two numbers gives the number of algebraic relations (`defining relations') which must persist among those products and multiple brackets in each stratum. These relations were explicitly calculated in laborious computer-aided calculations for $\mf h_m$ given  over alphabets of $3$ and $4$ letters up to stratum $\ell=7$ and $\ell=5$, respectively. Also the action of $\mf a$ on $\mf U$ was calculated explicitly up those strata. Notice that in the concrete calculations,  the auxiliary Lie algebra $\mf g$ was used.

The (modified) exceptional elements and the standard invariants from $\mf h_m^0$ and $\mf h_m^1$  are all of homogeneity degree 2. The conjecture that $\mf h_m$ is generated by them, has lead to a more general hypothesis, the  {\bf quadratic generation hypothesis}:  It is conjectured that the Poisson algebra $\mf h$ (not only the rest frame algebra $\mf h_m$) is (infinitely and not freely) generated as a Poisson algebra by invariants of homogeneity degree 2. \label{hyp:quad}\newline

\vspace{-2ex}
Regarding the quantization of $\mf h$, our discussion from the beginning of this section, although not directly applicable, justifies that a quantization of $\mf h_m$ should be given by a filtered associative algebra $\hat{\mf h}_m$ with multiplication of degree +1 and commutator of degree 0, such that for each filtration degree $\ell$, we have $\mf h_m^\ell \cong \hat{\mf h}_m^\ell/\hat{\mf h}_m^{\ell-1}$ as vectorspaces, and Poison brackets in $\mf h_m$ are reproduced modulo a deformation parameter $h$ (quantization conditions). The original Pohlmeyer programme is to construct this quantization {\em from $\mf h_m$ alone}, i.e. without reference to other algebras such as $H$.  See~\cite{pm_quant} and for a discussion, also~\cite{meusb_diplom,rm_real}. Essentially, the idea is to freely generate a filtered associative algebra from abstract  generators assigned to each generator of $\mf h_m$, and then to find an ideal in this algebra such that the  quotient with respect to this ideal satisfies the quantization conditions. The ideal is constructed stratum per stratum in the filtration, by deforming the defining relations in $\mf h_m$ (and the action of $\mf a$ on $\mf U$) by admitting quantum corrections, and the consistency of  this deformation has to be checked stratum per stratum in complicated computer-aided calculations. 

The obvious weakness of this approach is  that, unless one finds an obstruction to the consistency requirements, it cannot decide the question whether a quantization 
actually exists, since there is an infinite number of conditions that have to be checked one after the other (stratum per stratum). On the other hand, the calculations performed over the years did help in forming conjectures on the algebra's structure (or in disproving them).  For example, it could be shown that the semidirect product structure of the classical algebra is destroyed by the deformation: exceptional elements in general appear as quantum corrections in deformed relations.
This has in fact lead to the very interesting conjecture that -- contrary to the Poisson algebra $\mf h$ -- the quantum algebra $\hat{\mf h}_m$ might be finitely generated.


\subsection{A quantization procedure based on $\mf g$}\label{subsec:rm}

A major step forward in quantizing $\mf h$ was then achieved by Meusburger and Rehren in~\cite{rm_real}. Their approach is based on the following observation which we formulate in terms of our general remarks from the beginning  of this section:

{\rem \label{rem:AquantH} {\rm We first observe that  the symmetric enveloping algebra $S(\mf g)$ of the auxiliary Lie algebra $\mf g=\ima \, e$ with the natural Poisson structure and the Shuffle Hopf algebra $(H,\#, \{,\}_g)$ are isomorphic as {\em graded} Poisson algebras. To see this, recall here that $H$ is endowed with the combined degree  (\ref{eq:combgrad}), i.e. for a monomial $e(X_1) \# \cdots \# e(X_k)\in H$ we have $\ell=n-k-1$  with $n=\sum_i |X_i|$, and, as discussed above, $S(\mf g)$  is endowed with the grading given by  $\ell= \sum \deg(e(X_i))+k-1$ for a monomial $e(X_1)\cdots  e(X_k) \in S(\mf g)$, so the claim follows since   $ \deg(e(X_i))= |X_i| -2$.
It follows immediately that the algebra $\mc A_{\mf g}$ given in (\ref{eq:QuantLie}) 
is a quantization of the Poisson algebra $(H,\#, \{,\}_g)$.

\
}}

Now, by Proposition~\ref{prop:hsubalg}, the physical observables form a Poisson subalgebra $\mf h$ in $H$. Of course, it is straightforward to assign to each observable a lifting in $\mc A_{\mf g}$ by (\ref{eq:lifting}) with some quantum corrections which are themselves liftings of observables. However, it {\em cannot} be guaranteed that such a prescription will lead to a  subalgebra in $\mc A_{\mf g}$ whose elements are liftings of observables: taking a commutator of the liftings of two elements of $\mf h$, quantum corrections might appear which are not themselves liftings of observables.

In principle, one might try to fix the quantum corrections  by hand for a set of generators of the Poisson algebra of observables, $\mf h$, in such a manner that their commutators only produce liftings of observables. However, due to the fact that we do not have a set of {\em free} generators, cf.~the previous section~\ref{subsec:originalPM}, this would again amount to a never-ending calculation to be performed stratum per stratum in the filtration of $U(\mf g_h)$. Therefore, the following strategy was proposed in~\cite{rm_real}:

{\rem
\label{rem:rm_qcorr}
{\rm (Meusburger-Rehren approach):  For a set of generators of  $\mf h$, consider  liftings in $\mc A_{\mf g}$ of the form (\ref{eq:lifting}) with the property that their quantum corrections only occur from reordering factors in the leading order contribution (which, by construction, is a polynomial in $U(\mf g_h)$). Find a derivation of $U(\mf g_h)$, such that these liftings are in the kernel of this derivation's extension to $\mc A_{\mf g}$, and such that these liftings generate the full kernel. It then follows automatically that every element of $\mf h$ has a lifting in the kernel and that every element of the kernel is a lifting of an element in $\mf h$. Since the kernel of a derivation is a subalgebra, the latter implies, in particular, that all quantum corrections which can appear by taking commutators of elements of the kernel are again liftings of observables.

\
 }} 

Unfortunately, such a derivation has not yet been found in full generality. The difficulty is that one takes  a {\em Lie derivative} of $\mf g_h$ as a starting point, which then has a natural extension to a derivation of  $U(\mf g_h)$ -- but when asking for the map to be a Lie derivative of $\mf g_h$, one loses control over the kernel in $U(\mf g_h)$.  The authors of~\cite{rm_real} did succeed, however, in giving a derivation whose kernel is by construction mapped into $\mf h$ by the projection map, and which, moreover,  contains  liftings of invariants of homogeneity degree $2$. Provided that  the quadratic generation hypothesis holds, i.e. that the invariants of homogeneity degree $2$ generate all of $\mf h$, 
the projection map is also surjective, so the prerequisites listed in the above remark are satisfied.
So, under the quadratic generation hypothesis, this provides a consistent quantization of $\mf h$ as a subalgebra in $\mc A_{\mf g}$. This idea will be stated in Corollary~\ref{cor:mainthm} below. Observe that the construction is entirely independent of the size of the alphabet $A$ (no critical dimension is needed).

\medskip We will now recount the construction of this derivation from~\cite{rm_real} in the framework explained in the beginning of this section and the language from section~\ref{sec:euler}. We will also point out why it is so difficult to find a derivation that does not need the quadratic generation hypothesis.

As a first attempt to construct the derivation, one might think of the derivation $\partial$ of the Shuffle algebra along the morphism $\varphi(x)=1\otimes x$ which we encountered in the proof of Proposition~\ref{prop:hsubalg}, cf. (\ref{eq:partial}),
\[
\partial:H\rightarrow H \otimes H \, , \ \partial(x_1\cdots x_n)=x_1 \otimes x_2 \cdots x_n - x_n\otimes x_1\cdots x_{n-1}  \, , \ \partial(x_1)=0 \ \mbox{ for } x_i \in A
\]
\nopagebreak and whose kernel is $\mf h$. We first note:

{\lem \label{lem:partialime}{\rm Restricted to $\ima\, e$, the derivation $\partial$ takes the following form:
\beq\label{eq:partialime}
\partial(e(x_1\cdots x_n))=e(x_1)\otimes e(x_2 \cdots x_n) - e(x_n)\otimes e(x_1\cdots x_{n-1})  \, , \ \partial(e(x_1))=0 \ \mbox{ for } x_i \in A\, ,
\eeq
so, using $e(a)=a$ for $a \in A$, we have  $\ima\; \partial|_{\ima e} = H_1\otimes \ima \, e \subset \ima\,  e \otimes  \ima\, e \subset H \otimes \ima\, e$.
}}


The proof is included in Appendix~\ref{app:partial}, cf. also~\cite{meinecke_diplom}.
In fact, this was the form in which the derivation was defined in~\cite{rm_real} without explicitly giving its form on the full word algebra. From (\ref{eq:partialime}) alone, however, it is difficult  to see that after a suitable extension to $H$, the kernel of $\partial$ is indeed is $\mf h$. So, this was proved by an indirect argument in~\cite{rm_real} instead. 

Now, if $\partial$ could be understood as a derivation of Lie algebras, by putting an appropriate Lie bracket on $H \otimes \ima\, e$, then it could be extended to $U(\mf g_h)$ and would automatically have the correct kernel. We show, however, that this is not possible in the following sense:

{\prop \label{prop:partialnoLie}{\rm There is no  Lie bracket $[[,]]$ on the image of $e$, such that $\partial: \mf g \rightarrow H \otimes \ima \, e$ is a derivation of Lie algebras, where $H \otimes \ima \, e$ carries the natural Lie structure
\[
[[ x \otimes e(y), x^\prime \otimes e(y^\prime) ]] = x \# x^\prime \otimes [[e(y),e(y^\prime)]]  
\]
and $e(y) \in \ima \, e$ is identified with $1\otimes e(y) \in H \otimes \ima \, e$. 
}}

To prove this, we do not have to specify the explicit form of the structure constants of $\mf g$. Once the claim is proved for $\mf g$ with arbitrary structure constants, it automatically follows for $\mf g_h$. Along this line of reasoning, we will, unless otherwise stated, understand claims made for $\mf g$ to be true also for $\mf g_h$.

{\proof {We calculate, for $a,b,c,d \in A$,
\beqa
\partial\big([e(ab),e(cd)]\big) &=&  
g_{bc}  \partial \big(e(ad)\big) -  g_{bd} \partial \big(e(ac)\big) - 
g_{ac}  \partial \big(e(bd)\big) +  g_{ad} \partial \big(e(bc)\big)  
\\&=&
  e(a) \otimes (g_{bc} e(d) -g_{bd} e(c)) - (g_{bc} e(d)-g_{bd} e(c))\otimes e(a) - (a\leftrightarrow b )
\eeqa
On the other hand, $[[\partial(e(ab)),1\otimes e(cd)]] + [[1\otimes e(ab), \partial(e(cd))]] $ is equal to
\[
e(a)\otimes  [[e(b),e(cd)]] -e(b)\otimes  [[e(a),e(cd)]]  + e(c) \otimes [[e(ab), e(d)]] - 
e(d) \otimes [[e(ab), e(c)]]
\]
so $[[,]]$ must satisfy
\[
[[e(b),e(cd)]]=g_{bc} e(d) -g_{bd} e(c) \qquad \mbox{ for all } b,c,d \in A
\]
At the same time, however, we must have 
\[
0 = \partial \big( [e(a),e(bcd)] \big)  \stackrel{!}{=}  [1\otimes e(a), \partial(e(bcd))]  =  
e(b) \otimes [[e(a),e(cd)]] - e(d) \otimes [[e(a),e(bc)]]
\]
for all $a,b,c,d \in A$ -- hence, a contradiction. \hfill $\square$}}

{\rem {\rm One might want to try to solve the contradiction in the proof by modifying the Lie algebra structure on $\mf g$ itself, i.e. change  the bracket of elements involving $e(a)$, $a \in A$ (in such a way that the Poisson bracket on $\mf h$ remains unchanged). However, the only non-trivial bracket which is compatible with the identity $e(x \# y) =0$ is
\beq\label{eq:emodbrack}
[e(a),e(x_1\cdots x_n)] = g_{a x_1} e(x_2 \cdots x_n) - g_{a x_n} e(x_1 \cdots x_{n-1}) 
\eeq
such that the Jacobi identity is violated, unless $n=2$ -- and this modification in itself does not suffice for our purposes. Moreover, this modification would change the Poisson bracket on $\mf h$.

\
}}

It is therefore desirable to modify  $\partial$ on $H$ instead, such that the kernel of the new map $\tilde \partial$ is still equal to the cyclically symmetrized linear combinations of words $\mf h$, but such that its restriction to $\mf g$ is a Lie-derivative in the sense of the proposition above or, possibly, such that it is a cocycle $\tilde \partial :\mf g \rightarrow \mf g \wedge \mf g$. Here, of course, $\mf g \wedge \mf g$ is the antisymmetric tensor product of $\mf g$  with itself as a $\mf g$-module with the adjoint action
\[
[e(x),e(y)\wedge e(z)] = [e(x),e(y)]\wedge e(z) + e(y)\wedge [e(x),e(z)]
\]
and the cocycle condition is
\[
\tilde \partial ([e(x),e(y)]) \stackrel{!}{=} [\tilde \partial e(x), 1 \otimes e(y) +e(y)\otimes 1]
+[1 \otimes e(x) +e(x)\otimes 1, \tilde \partial e(y)]
\]
We have tried a number of constructions, among them  the natural map to the antisymmetric tensor product 
$H \wedge H$
\[
\Delta - \tau \Delta \, :  H \rightarrow H \wedge H
\]
with the deconcatenation coproduct $\Delta$ and the flip map $\tau(x\otimes y)=y\otimes x$. On 
$\ima \,e (H_1)$ and $\ima \,e (H_2)$, this map coincides with $\partial$ and the same line of reasoning as in proposition~\ref{prop:partialnoLie} shows that it violates the cocycle condition. Observe also that this map  does not restrict nicely to $\ima \, e$: By explicit calculation, preferably using formula (\ref{eq:edesc}) from the appendix, one finds that for $a,b,c,d \in A$,
$(\Delta - \tau \Delta) (e(abcd))$ produces (among other contributions) the term $\tsf 1 3 \,  ab \otimes cd$ while $(\id -\tau)( e(a) \otimes e(bcd) +  e(ab) \otimes e(cd) + e(abc) \otimes e(d)$ contains this term with a coefficient $\frac 1 4$, so   $(\Delta - \tau \Delta)\, e \neq (\id -\tau) ( e \otimes e)  \Delta $.

Of course, there are cocycles $\mf g \rightarrow \mf g\wedge \mf g$, among them the 
maps
\beq\label{eq:lieBi}
\partial_{a,y}: e(x) \mapsto  e(a) \wedge [e(y),e(x)] 
\eeq
for $a \in A$ and $y \in H$ fixed, which  even turn $\mf g$ into a Lie bialgebra (see below). But we have not yet found a cocycle possessing an extension to $H$ with kernel equal to $\mf h$. We will comment on this again in the final section of this paper.

Meusburger and Rehren, on the other hand, succeeded in constructing  a Lie derivative using an extension of $\mf g$,  where the loss of control over the kernel of its extension to $H$ is still manageable. We start the discussion by recalling from~\cite{rm_real}: 

{\lem {\rm The map $\alpha: \mf g \times H_1 \rightarrow H_1$, $\alpha(e(x),c) =: e(x).c$ given by  
\[
e(x).c = \left\{ \begin{array}{ll} 0 & \mbox{ for } |x| \neq 2 
\\
g_{ac} b - g_{bc}a & \mbox{ for } x= ab, \ a,b \in A \end{array} \right.
\]
defines  an action of $\mf g$ on the vectorspace $H_1$.}}

For the proof see~\cite{meusb_diplom}. The compatibility with $e(x\#y)=0$, $x,y \neq 1$, is not difficult to check, so the proof mainly requires checking the condition $[e(x),e(y)] . a = e(x). (e(y) . a)  - e(y) . (e(x) . a)$. For words which both have length unequal to $2$, this is clearly satisfied, so the non-trivial cases are those where one or both of them are of length 2.

Denote by $\tilde{\mf g}$ the corresponding semidirect product of $\mf g$ with the trivial Lie algebra $H_1$, i.e. the vector space $\mf g \times H_1$ with bracket 
\[
[(e(x),a),(e(y),b)] = ([e(x),e(y)], e(x).b -e(y).b)
\]

Inspired by~\cite{rm_real}, we now consider the following map:

{\prop \label{prop:delta0}{\rm The linear map $\delta_0: \mf g \rightarrow U(\tilde{\mf g}) $ given by 
\[
\delta_0(e(x_1\cdots x_n)) = \left\{ \begin{array}{ll}
v_{x_1} e(x_2\cdots x_n) - v_{x_n} e(x_1\cdots x_{n-1})  &\mbox{ for } n \geq 2
\\ 0 & \mbox{ otherwise } \end{array} \right. 
\]
is a derivation of Lie algebras. Here, $x_i \in A$ and in order to be able to distinguish the elements of $H_1$ from those of $\mf g$ of length 1, we have denote the basis elements in $H_1$ by $v_a$, $a \in A$.
}}

\medskip The proof of this claim is a straightforward, though lengthy calculation.
The important thing to note is that the non-trivial commutation relations of the $v_a$ with elements of $\mf g$ are  crucial to prove the derivation property, e.g. we have
\[
\delta_0 ([e(ab),e(cd)])
=v_a  (g_{bc} e(d) -g_{bd} e(c)) - (g_{bc} v_c-g_{bd} v_c) \, e(a) - (a\leftrightarrow b )
\]
while $[\delta_0 (e(ab)),e(cd)] + [e(ab),\delta_0 (e(cd))]$ is equal to 
\[
=[v_a,  e(cd) ] \, e(b)  - [v_b, e(cd) ]\, e(a)  + [e(ab), v_c ] \,e(d)
- [e(ab),  v_d ]\,e(c)
\]
This line of reasoning also shows that $\delta_0$ cannot be extended to a Lie derivative $\tilde{\mf g} \rightarrow U(\tilde{\mf g})$.

Now, since $\delta_0 \neq \partial$, we have to check whether there is an extension of $\delta_0$ to $H$ (by the Leibniz rule with respect to the shuffle multiplication) which still has $\mf h$ as its kernel. Since the shuffle multiplication is commutative, the only way in which $\delta_0$ can be extended is
\beq\label{eq:extdelta0}
\delta_0(e(X_1) \# \cdots \# e(X_k))= \tsf 1 {k!}\sum_{\pi\in S_k} \sum_{i=1}^k e(X_{\pi(1)}) \cdots  \delta_0(e(X_{\pi(i)})) \cdots e(X_{\pi(k)})
\eeq
for words $X_1, \dots, X_k$. Observe, however, that $\delta_0$ cannot be a derivation of the Poisson bracket on~$H$. Now, a lengthy calculation, cf.~\cite{meinecke_diplom}, shows that
\beq\label{eq:obstrZ2}
\delta_0(Z(e^{(2)}(x_1\cdots x_n))) = \tsf 1 2 \; Z\big( (g_{x_n x_1} v_{x_2} - g_{x_n x_2}v_{x_1} + g_{ x_1 x_2} v_{x_n} - g_{x_n x_2}v_{x_1})\, e(x_3 \cdots x_{n-1}) \big)
\eeq
where the cyclic symmetrization map $Z$ acts on all letters in the expression. Therefore,
\[
Z(e^{(2)}(x_1\cdots x_n)) \notin \ker \delta_0 \subset H 
\]
Consider therefore the following modified derivation which was first given in~\cite{rm_real}:

{\prop\label{prop:delta} {\rm The linear map $\delta: \mf g \rightarrow U(\tilde{\mf g}) $ given by 
\beq\label{eq:delta}
\delta(e(x_1\cdots x_n)) = \left\{ \begin{array}{ll}
[v_{x_1}, e(x_2\cdots x_n)]_+ - [v_{x_n}, e(x_1\cdots x_{n-1})]_+  &\mbox{ for } n \geq 2
\\ 0 & \mbox{ otherwise } \end{array} \right. 
\eeq
where $[,]_+$ denotes the anticommutator, $[x,y]_+=\frac 1 2 (xy+yx)$, is a derivation of Lie algebras. Moreover,  $Z(e^{(2)}(x))$ is in the kernel of its extension to $H$ by (\ref{eq:extdelta0}) for any $x \in H$.
}}

\proof{The first statement follows from Proposition~\ref{prop:delta0} and the second by inspection of~(\ref{eq:obstrZ2}).}

{\rem{\rm Proposition~\ref{prop:delta} was first proved in~\cite{meusb_diplom}, where, however, $\delta$ was not extended to the Shuffle algebra $H$, but directly to $U(\mf g_h)$. The lifting of invariants of homogeneity degree~2 did not include any quantum corrections, see (\ref{eq:Z2lift}) below, so the calculations from~\cite{meusb_diplom} 
directly correspond to those we performed in $H$ to calculate (\ref{eq:obstrZ2}). We have chosen to first extend $\delta$ to the classical algebra $H$ in order to clarify that the problem is not related to the quantization.}}

\medskip In~\cite{meusb_diplom} it was moreover shown by explicit calculation that $Z(e^{(3)}(001101))$ is not in the kernel of the extension of $\delta$ to $U(\mf g_h)$. In spirit, we can again use these calculations to conclude that the invariants of higher homogeneity degree $K\geq 3$ are in general not in the kernel of the extension of $\delta$ to $H$. So far, this problem could not be solved  by a further modification of the map $\delta$, so we have not been able to extend the main statement from~\cite{rm_real}: 

{\cor \label{cor:mainthm} {\rm Let,  as explained in the beginning of this subsection, $\mc A_{\mf g}$ denote the quantization of the Poisson algebra $(H,\#,\{,\}_g)$ given by (\ref{eq:QuantLie}) with the projection map $\varphi:\mc A_{\mf g} \rightarrow H$ as in  (\ref{eq:varphi}) and liftings of the form (\ref{eq:lifting}).
Consider the natural extension $\delta: U(\mf g_h) \rightarrow U( \tilde{\mf g}_h)$ of the Lie derivation  $\delta : \mf g_h \rightarrow U( \tilde{\mf g}_h)$ given by (\ref{eq:delta}),
\[
\delta(e(x_1\cdots x_n)) = \left\{ \begin{array}{ll}
[v_{x_1}, e(x_2\cdots x_n)]_+ - [v_{x_n}, e(x_1\cdots x_{n-1})]_+  &\mbox{ for } n \geq 2
\\ 0 & \mbox{ otherwise } \end{array} \right. 
\]
with the anticommutator $[,]_+$. The extension's kernel is a filtered subalgebra in $U(\mf g_h)$ with strata $U(\mf g_h)_\ell\cap \ker \delta$.
By Lemma~\ref{lem:partialime}, $\varphi$ maps the corresponding kernel of $\delta$ in $\mc A_{\mf g}$ into the kernel of $\partial:H\rightarrow H \otimes H$ given by (\ref{eq:partial}), 
$
\partial(x_1\cdots x_n)=x_1 \otimes x_2 \cdots x_n - x_n\otimes x_1\cdots x_{n-1}$ for $n\geq 2$ and $\partial(x_1)=0 $ for letters $x_i \in A$, and since this kernel is, in turn, equal to the algebra of observables $\mf h$ (the cyclically symmetrized linear combinations of words), $\varphi|_{\ker \delta}$ takes values in $\mf h$.

Now assign to each element $Z(e^{(2)}(x_1\cdots x_n)) \in \mf h^{n-3}$ 
a lifting of the form 
\beq\label{eq:Z2lift}
\hat Z^{(2)} (x_1\cdots x_n) := h^{n-3}\,Z \big(\,  \tsf 1 2 \sum_{i=1}^{n-1} e(x_1 \cdots x_i) \, e(x_{i+1} \cdots x_n) \big) \in h^{n-3}U(\mf g_h)_{n-3}
\eeq
with the cyclic symmetrization map $Z$ acting on all letters. Then, by Proposition~\ref{prop:delta}, we have $\hat Z^{(2)}(x) \in \ker\, \delta \subset \mc A_{\mf g}$ for any word $x$.
It now follows that if the quadratic generation hypothesis is true, $\varphi$ maps the kernel of $\delta$ surjectively to  $\mf h$. Therefore, under this prerequisite, $\ker\, \delta$ is a quantization of $\mf h$.}}

\medskip {\remnn If the quadratic generation hypothesis should be false, not all invariants will have a quantum counterpart in $\ker \delta$ as in the statement above (cf. the example $Z(e^{(3)}(001101))$ again). Moreover, in this case, $\ker \delta$ would {\em not}, unfortunately, yield a quantization of the subalgebra $\mf h^\prime$ of $\mf h$ generated by invariants of homogeneity degree~2. The reason is that quantum corrections which appear here, will again be in the kernel of $\delta$ and therefore will be liftings of elements of $\mf h$, but there would be no guarantee in this case that they are liftings of elements of $\mf h^\prime$.}

\bigskip 

Let us conclude this section with some remarks on concrete calculations within this framework. Observe first that no quantum corrections appear in (\ref{eq:Z2lift}). However, such corrections will of course appear when we evaluate and reorder (multiple)  commutators  and products of quantum invariants $\hat Z^{(2)}(\cdots)$. 
In particular, liftings of invariants such as, say $Z(e^{(3)}(001101))$, are calculated in this setting by first writing the invariant  as a linear combination of (multiple) Poisson brackets and products of invariants $Z(e^{(2)}(\cdots))$, then replacing these invariants by their unique lifting (\ref{eq:Z2lift}), the brackets by $\frac 1 h$ times the commutator, and products by noncommutative products --  and then by admitting 
quantum corrections in the sense of (\ref{eq:lifting}) from the kernel of $\delta$. In~\cite{rm_real} these quantum corrections were  partly fixed by the requirement that they must all come from {\em reordering} the factors $e(\cdots)$ contributing to the leading order terms.
Applied to the defining relations of $\mf h$ (cf. section~\ref{subsec:originalPM}) this means the following: Such a relation is  given as a linear combination of (multiple) Poisson brackets and products of invariants $Z(e^{(2)}(\cdots))$ which add up to 0. In this linear combination, one  replaces the invariants by their unique lifting (\ref{eq:Z2lift}), the brackets by $\frac 1 h$ times the commutator, and products by anticommutators. This gives an element $X$ in $\mc A_{\mf g}$  which by our projection map  $\varphi$ is of course mapped to 0 in $\mf h$. Therefore, $X$ differs from 0 only by some reordering of its contributing factors $e(\cdots)$. The quantum corrections to the defining relation are then fixed by the requirement that, together with $X$, they add up to 0 in $\mc A_{\mf g}$. For a number of defining relations, the  quantum corrections have been calculated in this manner and are in agreement with those calculated within Pohlmeyer's original programme.


\section{Outlook} 
\setcounter{equation}{0}

We have seen that an approach to Pohlmeyer's Poisson algebra of invariant charges $\mf h$ in 
the framework of standard combinatorial algebra facilitates its  analysis, and that proposals for its quantization can be analysed within a general framework of deformation theory. We have also seen why it would be  very difficult to generalize the approach of Meusburger and Rehren in such a way that the quadratic generation hypothesis would no longer be needed. 

We moreover believe that using  the Hopf algebraic formulation, a detailed comparison with a more recent formulation of $\mf h$ as a Poisson algebra associated to a Quasi-Lie-bialgebra found by Bordemann, Enriquez and Hofer~\cite{beh_pm} is possible. Let us now sketch this approach as well as some open questions.
We first recall some definitions: A Quasi-Lie-bialgebra is a Lie algebra $\mf k$ equipped with a cocycle  $\delta: \mf k \rightarrow \mf k \wedge \mf k$ such that $cp(\delta \otimes \id) \delta : \mf k \rightarrow \mf k \wedge \mf k\wedge \mf k$ (where $cp(a\otimes b \otimes c)=a\otimes b \otimes c + c \otimes a \otimes b + b\otimes c \otimes a)$ is the coboundary of some $\varphi \in  \mf k \wedge \mf k\wedge \mf k$, i.e.
\[
cp(\delta \otimes \id) \delta (x) = [ cp(x \otimes 1 \otimes 1), \varphi]\ .
\]
Observe that a Quasi-Lie-bialgebra with $\varphi = 0$ is a Lie bialgebra. For example, the Lie algebra of Eulerian idempotents in the Shuffle algebra $\ima\, e=\mf g$ with cocycle $\delta_{a,y}$ given by (\ref{eq:lieBi}) is a Lie bialgebra. Consider now the universal enveloping algebra $U(\mf k)$, which is a Hopf algebra with coproduct given by the shuffle comultiplication $\Delta_\#$. Extend $\delta$ along $\Delta_\#$ to a derivation $D: U(\mf k) \rightarrow U(\mf k)\otimes U(\mf k)$. On the dual  $U(\mf k)^*$ consider the two operations
\beq\label{eq:assPoiss}
F \bullet G := (F \otimes G) \Delta_\# \mbox{ and } \{F,G\}_D:=(F\otimes G)D 
\eeq
Denote by ${\rm Tr}(U(\mf k))$ the set of all traces in $U(\mf k)^*$, i.e. all maps $F \in U(\mf k)^*$ with $F(ab)=F(ba)$. Then $({\rm Tr}(U(\mf k)),\bullet,\{,\}_D)$ is a Poisson algebra, the Poisson algebra associated to the quasi Lie Bialgebra $(\mf k, \delta, \varphi)$.

The important observation in~\cite{beh_pm} was that Pohlmeyer's Poisson algebra $\mf h$ of invariants can be understood as a Poisson algebra associated to a Quasi-Lie-bialgebra. To do so, the authors first considered the iterated integrals (\ref{eq:Rpm}) and (\ref{eq:Zpm}) from the Introduction  as maps 
\[
\mc R^\pm: T \mathbb M \rightarrow {\rm Fun}(\mc P, C^\infty(\Sp^1,\R))\quad \mbox{ and } \quad
\mc Z^\pm: {\rm Tc}(T(V)) \rightarrow {\rm Fun}(\mc P, \R)
\]
where $\mathbb M = \R^{1,d-1}$ is the Minkowski space, $V$ its dual $\mathbb M^*$, $\mc P=C^\infty(\Sp^1, \mathbb M \times \mathbb M)$ is the system's phase space,
and where ${\rm Tc}(T(V))$ denotes the space of restricted traces ${\rm Tr}(T(V)) \cap T \mathbb M$ (which is isomorphic to the space of cyclic tensors in $T\mathbb M$, hence the span of the invariant charges $\mf h$). They then proved:

{\prop \label{prop:beh} {\rm  Let $L(\mathbb M^*)$ denote the free Lie algebra over $\mathbb M^*$, let $\eta \in S^2(\mathbb M^*)$ be the Minkowski metric on $\mathbb M$. Then $(L(\mathbb M^*),\delta_\eta,\varphi_\eta)$ is a Quasi-Lie-bialgebra, where 
\[
\delta_\eta (x) = [\eta,x\otimes 1 -1 \otimes x] \mbox { and } \varphi_\eta =  - [\eta^{12},\eta^{13}-\eta^{23}] - [\eta^{13},\eta^{23}]
\]
in the usual Hopf algebra notation, e.g. 
 $\eta^{13}=\sum_i \eta_{1,i} \otimes 1 \otimes \eta_{2,i}$ for $\eta=\sum_i \eta_{1,i}\otimes \eta_{2,i}$. 

The space of restricted traces ${\rm Tc}(T(\mathbb M^*))$ forms a Poisson algebra $({\rm Tc}(T(\mathbb M^*), \bullet, \{,\}_D)$) with product and Poisson bracket given by (\ref{eq:assPoiss}), and we have for $V,U \in {\rm Tc}(T(\mathbb M^*))$,
\[
\langle \mc Z^\pm ,U\rangle\,\langle \mc Z^\pm ,V\rangle = \langle \mc Z^\pm ,U \bullet V \rangle
\ \mbox{ and } \ \{ \langle \mc Z^\pm ,U\rangle\, , \, \langle \mc Z^\pm ,V\rangle \}_g = \langle \mc Z^\pm , \{U , V \}_D \rangle
\]
with the natural pairing $\langle, \rangle$ and the Poisson bracket $\{,\}_g$ given by (\ref{eq:PoisBrZ}), with $g=-2\eta$. }}

\bigskip Observe on the other hand, that explicit calculations show that the Poisson algebra associated to the Lie bialgebra $\mf g = \ima \, e$ with coclucle $\delta_{a,y})$ given in  (\ref{eq:lieBi}) is not isomorphic to $\mf h$ for any choice of words $y$ and letters~$a$. An interesting open question is, however, whether there is a structural connection between the existence of an underlying Quasi-Lie-bialgebra structure for $\mf h$ and the existence of the Lie structure on $\ima \, e$ -- and whether such a connection might be present in a more general context.

The quantization of Quasi-Lie-bialgebras is by now understood~\cite{enriquez}. By Proposition~\ref{prop:beh}, it guarantees the existence of a quantization of $\mf h$ -- which, contrary to the approach of Meusburger and Rehren, does not rely on the still open conjecture on the quadratic generation property.  Note that explicit calculations from the earlier paper~\cite{beh_pm} performed for the  coboundary case indicate that this quantization differs from the explicit realization proposed by Meusburger and Rehren. A thorough understanding of this matter is related to the question raised above, and should be pursued as well.

From the point of view of physics, the difficult question of constructing (Hilbert space) representations of any quantization of $\mf h$ is of great importance. Provided that the quadratic generation property can be proved, it seems that the approach based on the auxiliary Lie algebra $\mf g$ is promising: An ongoing analysis\footnote{D. Bahns and N. Hansen, work in progress.} of its precise structure (at least for alphabets of length $3$ and $4$) indicates that it possesses a root space decomposition that might be helpful in this respect. Although it cannot be excluded that a restriction on the dimension of the embedding vector space to some critical dimension might appear when representations are studied, this is not to be expected, especially in view of the results from~\cite{bahns_CQ}. We therefore  point out again that in none of the  approaches to quantizing $\mf h$ explained above such a restriction is necessary.


\bigskip {\bf Acknowledgement} 

Dorothea Bahns would like to thank Martin Bordemann for a very helpful discussion in Oberwolfach in 2010.

\medskip 

Supported by the German Research Foundation 
(Deutsche Forschungsgemeinschaft (DFG)) through
the Institutional Strategy of the University of G\"ottingen.


\begin{appendix}

\section{Proof of Lemma~\ref{lem:partialime}}\label{app:partial}
\setcounter{equation}{0}
\renewcommand{\theequation}{A.\arabic{equation}}

We first note that in the Shuffle algebra, $e(x_1\cdots x_n)$, $x_i \in A$, can be written as a linear combination of words given by all permutations $\pi \in S_n$ of the letters $x_1, \dots, x_n$ with coefficients given in terms of $d_\pi$, the number of descents in $\pi$, i.e. the number of $i$ for which $\pi^{-1}(i)>\pi^{-1}(i+1)$, as follows
\beq\label{eq:edesc}
e(x_1 \cdots x_n) =\sum_{\pi\in S_n} c(n,d_\pi) \ x_{\pi(1)} \cdots x_{\pi(n)} \qquad \mbox{ with } c(n,d_\pi)  = \frac {(-1)^{d_\pi}} n \left({n -1} \atop {d_\pi}\right)^{-1}
\eeq
For a proof see~\cite{lod_idem} (in the dual algebra) and also~\cite{meinecke_diplom}.

Now  consider $\partial(e(x_1\cdots x_n)) = \sum_{\pi\in S_n} c(n,d_\pi) \ \partial (x_{\pi(1)} \cdots x_{\pi(n)})$, which is equal to
\beq
\sum_{\pi\in S_n} c(n,d_\pi) \ \big( x_{\pi(1)} \otimes x_{\pi(2)} \cdots x_{\pi(n)} - x_{\pi(n)}\otimes  x_{\pi(1)} \cdots x_{\pi(n-1)} \big) \, , \label{A:partiale}
\eeq
and  observe that (\ref{A:partiale}) can be split  into contributions of the following form, 
\beqan
&&\sum_{\sigma \in S_{n-1}} k_\sigma \ x_1 \otimes x_{\sigma(2)} \cdots x_{\sigma(n)}\label{1e}\\
&+&\sum_{\sigma \in S_{n-1}} \sum_{i=2}^{n-1} l_\sigma \ x_i \otimes x_{\sigma(1)} \cdots x_{\sigma(i-1)}x_{\sigma(i+1)} \cdots x_{\sigma(n)}\\
&+&\sum_{\sigma \in S_{n-1}} k^\prime_\sigma \ x_n \otimes x_{\sigma(1)} \cdots x_{\sigma(n-1)}\label{ne}
\eeqan
with as yet unknown coefficients $k_\sigma, l_\sigma, k^\prime_\sigma$. We will now show that $l_\sigma=0$ while $k_\sigma$ and $k^\prime_\sigma$ are in fact $c(n-1,d_\sigma)$. This will prove the claim.

The terms in line~(\ref{1e}) are produced firstly from $x_{\pi(1)} \otimes x_{\pi(2)} \cdots x_{\pi(n)}$ by all permutations $\pi$ of the form $\pi(1)=1$ and $\pi(i)=\sigma(i)$, $i\geq 2$, and secondly, from 
$ - x_{\pi(n)}\otimes  x_{\pi(1)} \cdots x_{\pi(n-1)}$ by all permutations $\pi^\prime$ of the form $\pi^\prime(n)=1$ and $\pi^\prime(i-1)=\sigma(i)$, $i\leq n$. Therefore, we have $k_\sigma=c(n,d_{\pi})-c(n,d_{\pi^\prime})$.
Now, the number of descents of $\pi$ is the same as that of $\sigma$, while  $\pi^\prime$ has one descent more, so 
\[
k_\sigma = \frac {(-1)^{d_\pi}} n \left({n -1} \atop {d_\pi}\right)^{-1}
  -\  \frac {(-1)^{d_\pi+1}} n \left({n -1} \atop {d_\pi +1 }\right)^{-1} =  \ \frac {(-1)^{d_\pi}} {n-1} \left({n -2} \atop {d_\pi}\right)^{-1}  = \ c(n-1,d_\sigma)
\]
The same argument shows that the coefficient in (\ref{ne}) is $k^\prime_\sigma=c(n-1,d_\sigma)$.  And finally, a similar line of thought shows that the two permutations $\pi$ and $\pi^\prime$ which produce 
$x_i \otimes x_{\sigma(1)} \cdots x_{\sigma(i-1)}x_{\sigma(i+1)} \cdots x_{\sigma(n)}$ from 
$x_{\pi(1)} \otimes x_{\pi(2)} \cdots x_{\pi(n)}$ and 
$ - x_{\pi(n)}\otimes  x_{\pi(1)} \cdots x_{\pi(n-1)}$, respectively, have the {\em same} number of descents. Therefore, the difference of the corresponding coefficients, $l_\sigma = c(n,\pi)-c(n,\pi^\prime)$, is $0$.

\end{appendix}



\end{document}